\documentstyle[twoside,fleqn,espcrc2,epsf]{article}

\def\itmb{\begin{itemize}}
\def\itme{\end{itemize}}
\def\enmb{\begin{enumerate}}
\def\enme{\end{enumerate}}
\def\eqnb{\begin{equation}}
\def\eqne{\end{equation}}
\def\eqab{\begin{eqnarray}}
\def\eqae{\end{eqnarray}}
\def\dis{\displaystyle}
\def\NPB{{\em Nucl. Phys.} B}
\def\PLB{{\em Phys. Lett.}  B}

\newcommand{\AmS}{{\protect\the\textfont2
  A\kern-.1667em\lower.5ex\hbox{M}\kern-.125emS}}

\title{ Numerical studies of confinement in the lattice Landau gauge}

\author{Hideo Nakajima\address
{Department of Information Science, Utsunomiya University,\\
Utsunomiya 321-8585 Japan (e-mail nakajima@is.utsunomiya-u.ac.jp)},
Sadataka Furui\address
{School of Science and Engineering, Teikyo University, \\
Utsunomiya 320-8551, Japan (e-mail furui@liberal.umb.teikyo-u.ac.jp)}
and Azusa Yamaguchi\address
{Department of Physics, Ochanomizu University,\\
Tokyo 112-8610 Japan (e-mail: azusa@sokrates.phys.ocha.ac.jp).}
}

\begin{document}

\begin{abstract}
Critical conjectures on confinement in the Landau gauge is
numerically tested in focus to Gribov copy effects.
One of the subjects is of the Kugo-Ojima confinement criterion
and the other is of various viewpoints in the Gribov-Zwanziger theory.
We use the smearing gauge as a reference gauge free of Gribov copy, and
performed three types of simulations,
$\log U$, $U$-linear and $\log U$ in the smearing gauge.
It is found that Gribov copy effect on the Kugo-Ojima parameter is 
small. $\log U$ and $U$-linear simulations yield
only  global scale factor difference in gluon propagator and in 
ghost propagator,
and about 10\% difference in Kugo-Ojima parameter.
The horizon function defined by Zwanziger is evaluated in
three types of gauge field and compared. All data show the negative
horizon function as expected.
\vspace{1pc}
\end{abstract}

\maketitle

\section{INTRODUCTION}
Gribov pointed out that there exists a fundamental problem with
respect to the field theoretic description of QCD in its quantization,
due to gauge fixing degeneracy\cite{Gv}.
%This problem remains still unresolved as a generic and basic aspect of QCD
%other than in Landau or Coulomb gauges as well.
 Gribov conjectured that
the linear potential of quarks may be caused by an enhancement of the
singularity of the ghost propagator due to the restiriction of the gauge
field on the transverse plane. Zwanziger developed further the lattice gauge
theory along the line of Gribov, and discussed its continuum limit\cite{Zw}.
%in use of his restriction hypothesis of the core region .
\par
Kugo ond Ojima developed the operator formalism
of QCD starting from the standard BRST (Becchi-Rouet-Stora-Tyutin) symmetric
Lagrangian, and derived Kugo-Ojima criterion, a sufficient condition
for the colour confinement, i.e.,
the absence of free single coloured states in the asymptotic Hilbert
space\cite{KO}. Even though there is no question in validity of the
starting Lagrangian as the perturbation theory,
the validity is unclear as the full theory, if one thinks of the Gribov
problem which emerges in the 'large' amplitude region of the
gauge field. There is a discussion of Fujikawa\cite{Fuj} and
Hirschfeld\cite{Hir} trying to justify the standard Lagrangian.
This point is intimately related to a problem in numerical
measurement\cite{MO} of gauge non-invariant quantity in presence of
 Gribov copy\cite{NFY}.\par
The Kugo-Ojima theory and Gribov-Zwanziger theory
appear to stand on different footings, but
each presents its own critical conjectures on confinement problems.
It is meaningful to explore their validity numerically by lattice QCD
dynamics. We test how numerically satisfied is the Kugo-Ojima criterion
that a coefficient $u^a_b(0)$ in the two-point function produced by 
the ghost, the antighost and the gauge field becomes $-\delta_a^b$, 
and
investigate how gluon propagator and ghost propagator behave in the infrared
region. We also look at value of Zwanziger's horizon function to test
if his horizon condition is satisfied\cite{Zw}.
The theoretical problem of quantization caused by the Gribov copy
reflects as the Gribov noise problem in measurement of gauge non-invariant
quantity, and it is favoured to choose a unique copy among others,
i.e., {\bf a new gauge without Gribov copy} 
(see \cite{NFY} in more detail). As finding algorithms of unique gauge fixing
is difficult in various gauges in general, random statistical method is
commonly used, but it is cautioned that its performance is sensitive to
incorporated algorithms in some cases\cite{Bny}.  Laplacian gauge fixing is well known as a unique gauge fixing, but only known method in Landau gauge is
smearing gauge fixing\cite{HdF}. We call this gauge as smearing gauge
and adopt it as a reference gauge to investigate the Gribov copy effect.\par
There are some options of gauge field definition $A_\mu(U)$
in lattice gauge theory.
We compare numerical data with respect to two types of
definition, $\log U$ and $U$-linear.
\section{Gauge fields and lattice Landau gauge}
Options of $A_{\mu}(U)$ for $SU(3)$ lattice QCD are 
\begin{enumerate}
\item 
$\log U $;\ 
$U_{x,\mu}=e^{A_{x,\mu}},\ A_{x,\mu}^{\dag}=-A_{x,\mu},$
where $|$eigenvalue of $A_{x,\mu}|\le 4\pi/3\cite{NF}$.
\item
$U$-linear;\ 
$A_{x,\mu}=\displaystyle{1\over 2}(U_{x,\mu}-U_{x,\mu}^{\dag})|_{trl.}
\cite{Zw}.$
\end{enumerate}

In both cases the Landau gauge, $\partial A^g=0$, can be
 characterized in use
of optimizing functions $F_U(g)$ of $g$, such that $\delta F_U(g)=0$ for
any $\delta g$, respectively;($trl$ means the traceless part)
\begin{enumerate}
\item 
$F_U(g)=||A^g||^2=\sum_{x,\mu}{\rm tr}
 \left({{A^g}_{x,\mu}}^{\dag}A^g_{x,\mu}\right)$,
\item 
$F_U(g)=\sum_{x,\mu}\left (1- {1\over 3}{\rm Re}{\rm tr}U^g_{x,\mu}\right).$
\end{enumerate}
In both options, the variation
of the optimizing function, $F_U(g)$, under infinitesimal gauge transformation
$g^{-1}\delta g=\epsilon$, reads as
\eqnb
\Delta F_U(g)=-2\langle \partial A^g|\epsilon\rangle+
\langle \epsilon|-\partial { D(A^g)}|\epsilon\rangle+\cdots,
\label{NORM1}
\eqne
where $D_\mu(A)$ denotes the covariant derivative in each definition respectively; in short notations,\\
$U=U_{x,\mu}$, $A=A_{x,\mu}$, and 
$\partial_\mu \phi=\phi(x+\mu)-\phi(x)$,
$\bar \phi=\dis{1\over 2}\left(\phi(x+\mu)+\phi(x)\right)$,
\enmb
\item 
%in $A=\log U$ version,
\eqnb
D_{\mu}(A)\phi=S({\cal A})\partial_\mu \phi
+[A,\bar \phi],
\eqne
where ${\cal A}=adj_{A}=[A,\cdot]$,
$
S(x)=\dis{{x/2\over {\rm th}(x/2)}}.
$
\item
%in $U$-linear version,
\eqnb
D_{\mu}(U)\phi=\dis{1\over 2}\left.\left\{ \dis{U+U^\dag\over 2},\partial_{\mu}\phi\right\}\right |_{trl.}+[A,\bar \phi].
\eqne
\enme
\section{The Kugo-Ojima confinement criterion and the Gribov-Zwanziger's theory}\subsection{The Kugo-Ojima criterion\cite{KO}}
The Kugo-Ojima criterion is stated as follows; $u^{ab}(0)=-\delta^{ab}$,
where $u^{ab}$ is defined by the following two-point
function as,
\[
(\delta_{\mu\nu}-{p_\mu p_\nu\over p^2})u^{ab}(p^2)
\]
\begin{equation}
=\displaystyle{1\over V}
\sum_{x,y} e^{-ip(x-y)}\langle  {\rm tr}\left({\lambda^a}^{\dag}
D_\mu \displaystyle{1\over -\partial D}[A_\nu,\lambda^b] \right)_{xy}\rangle
\eqne
where angle brackets denote sample average, and $\lambda^a$ is a normalized 
$({\rm tr}{\lambda^a}^\dag\lambda^b
=\delta^{ab})$ antihermitian basis of Lie algebra, $V$ a lattice volume.

\subsection{Zwanziger's theory\cite{Zw}}
Zwanziger classified transverse space of gauge field into some sets.
 $F_U(g)$ takes the local minimum in the Gribov region $\Omega$, and the
global minimum in the {\bf fundamental modular region} $\Lambda$,
$\Lambda\subset \Omega$. Within $\Lambda$, there exists no
gauge fixing degeneracy up to global gauge transformation. We call this
gauge $\Lambda$ gauge.
For purpose of gauging away Bloch wave states,
Zwanziger narrowed down $\Lambda$ to {\bf core region} $\Xi\subset\Lambda$
:
Let the space of configuration of period $L$ be $\Pi_L$, and the fundamental
modular region in the $\Pi_L$ be $\Lambda_L$, and let
$\Lambda_L^N=\Lambda_{NL}\cap \Pi_L$, then the core region is given by
$
\Xi_L\equiv \Lambda_L^\infty 
$. Note, however, that by this definition, considerably large amount of points
in $\Lambda_L$ fail to come into $\Lambda_L^N$ not because they
can be gauge transformed to the point in $\Lambda_L^N$, but because
they fail to satisfy $L$-periodicity condition when
gauge transformed to $\Lambda_{NL}$. From this view point, narrowing down
to the core region looks like a kind of artificial selection of configurations.
Thus Zwanziger's basic hypothesis is that partition
functions defined by path integrals over $\Xi_L$
and $\Lambda_L$, give the same limit, 
%\eqnb
$
\lim_{L\to \infty}Z_{\Xi_L}=\lim_{L\to\infty}Z_{\Lambda_L}.
%\eqne
$
Since this hypothesis is highly dynamical, it is to be
checked somehow, e.g., by simulation. One can show that
the following {\bf horizon function} $H(U)$ is negative
in the above defined core region $\Xi$ by estimating increase of $F_U(g)$
by Bloch wave excitation. $H(U)$ is given as follows;
\begin{equation}
H(U)=\sum_{x,y,a}{G_{\mu\mu xy}}^{aa}-8E(U),
\end{equation}
\begin{equation}
{G_{\mu\nu xy}}^{ab}
=
 {\rm tr}\left({\lambda^a}^{\dag}
D_\mu \displaystyle{1\over -\partial D}(-D_\nu)\lambda^b\right)_{xy},
\label{eqq3}
\end{equation}
with
\enmb
\item 
\eqnb
E(U)=\dis{{1\over 8}\sum_{l,a}}{\rm tr}\left ({\lambda^a}^\dag S({\cal A}_l)\lambda^a\right),
\eqne
\item
\begin{equation}
E(U)=\sum_l{1\over 3}{\rm Re}\ {\rm tr} U_l,
\end{equation}
\enme
Statistical average defines a tensor $G_{\mu\nu xy}$ as
$\langle G_{\mu\nu xy}^{ab}\rangle=G_{\mu\nu xy}\delta^{ab}$,
provided colour symmetry is not broken.
The Fourier transform of the tensor $G_{\mu\nu xy}$ 
takes a form
\begin{equation}
G_{\mu\nu}(p)\delta^{ab}=
\left(\dis{e\over 4}\right)\displaystyle{p_\mu p_\nu\over p^2}\delta^{ab}
-\left(\delta_{\mu\nu}-\displaystyle{p_\mu p_\nu\over p^2}
\right)u^{ab},
\label{GMN}
\end{equation}
where $e=\langle E(U)\rangle/V$.
It is related with the horizon function as
\begin{equation}
\displaystyle{\langle H(U)\rangle\over V}
=8\left[\lim_{p\to 0}
G_{\mu\mu}(p)-e\right].
\label{INFZW}
\end{equation}
Note that $e=4$ if all link $U_l=1$.
For purpose of path integral estimation of the partition function, a set
{\bf augmented core region} $\Psi=\{U : H(U)\le 0\}\cap \Omega$
 ( $\Xi\subset\Psi\subset \Omega$ ) is defined, and from estimation of
$Z_{\Psi}$, one
derives the infinite volume limit\cite{Zw} that
$\lim_{V\to \infty}\displaystyle{\langle H(U)\rangle\over V}=0$
which is called horizon condition. Putting Kugo-Ojima parameter as
$
u^{ab}(0)=-\delta^{ab}c
$, one finds from (\ref{GMN}), (\ref{INFZW}), that the horizon condition is
written as 
$
\left(\dis{e\over 4}\right)+3c-e=3\left(c-\dis{e\over 4}\right)=0.
$

\begin{table*}[htb]
\caption{Results of the Kugo-Ojima parameter $c$, 
 'trace' $e$ divided by the dimension 4, and $h=c-e/4$.}
\label{table:1}
\newcommand{\m}{\hphantom{$-$}}
\newcommand{\cc}[1]{\multicolumn{1}{c}{#1}}
\renewcommand{\tabcolsep}{2pc} % enlarge column spacing
\renewcommand{\arraystretch}{1.2} % enlarge line spacing
\begin{center}
\begin{tabular}{@{}llll}
\hline
gauge fixing           & {$c$} & {$e/4$} & $h$  \\
\hline
$\log U$           & 0.628(94)  & 0.943(1)  & -0.32 \\
smeared $\log U$   & 0.647(101) & 0.943(1)  & -0.30 \\
$U$-linear         & 0.576(79)  & 0.860(1)  & -0.28 \\
\hline
\end{tabular}%\\[2pt]
\end{center}
\end{table*}

\section{Method of Landau gauge fixing, numerical results and discussion}
Our simulation was done for $\beta=6.0$ in lattice size $16^4$.
Simulation of gauge fixed theory can be simply executed by gauge transforming
a Boltzmann sample to the gauge, and measuring a quantity in question\cite{MO}.
 This algorithm is unambiguous and correct only if the gauge is
free of Gribov copy, or the quantity is gauge invariant.
It is easily understood that
if the quantity is gauge non-invariant and if there exists Gribov copy,
numerical results depend not only on the gauge itself, but also on the
algorithm employed which brings a sample to a gauge copy\cite{Bny}.
We performed three types of Landau gauge calculations, 
combinations of gauge field definitions and gauge fixing algorithms.
\enmb
\item
Gauge fixing with $\log U$ option using a direct Newton method, 
i.e., gauge transformation phase $\phi$ being given by
$\phi=(-\partial D(A))^{-1}\partial A$\cite{NF}. Gribov copy exists.
\item
Gauge fixing with $\log U$ option using the smearing gauge fixing\cite{HdF}.
This smearing gauge is a unique Landau gauge. But our execution shows
that the smearing gauge at $\beta=6$, $16^4$ does not coincide
completely with $\Lambda$ gauge.
\item
Gauge fixing with $U$-linear option using the non-stochastic-over-relaxation
method after the process 1. Gribov copy exists.
\enme
 
In our smearing gauge fixing, accuracy of plaquette value $E_p$ of 
smeared vacuum configuration is set as $1-E_p<10^{-7}$,
since there appear a long plateau 
around $1-E_p\sim 10^{-4}$ in smearing process for many samples,
and the finally obtained gauge fixed configuration 
depends critically on the situation if the smearing is finished before
or after the long plateau.

Kugo-Ojima parameter is found to be $u^a_b(0)\sim -0.65$ for cases 1,2.
and there is no remarkable difference. The magnitude of the parameter is
about 10\% less in case 3. 
$u^a_b$ does not look like tending to -1
up to the lattice size $16^4$. The reason is unclear so far.
And Table 1 shows the Kugo-Ojima parameters
and values related to the horizon function. All values indicate
negative horizon function\cite{STR}.

Gluon propagator is infrared finite, and $\log-\log$
plot of unrenormalized gluon propagator is given in Fig 1, and about 20\%
 difference of
case 3 appear as a global scale factor, i.e., wave function
renormalization, and this situation
agrees with the discussion in the literature\cite{Gius}. Case 2 is almost same
as case 1.

\begin{figure}[htb]
\begin{center}
\epsfysize=110pt\epsfbox{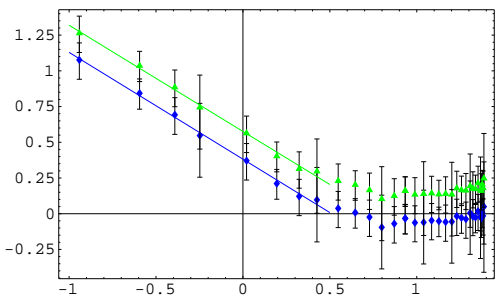}
\caption{The $\log$ of the unrenormalized gluon propagator multiplied by $(qa)^2$ as a function of the $\log qa$ where $q$ is the lattice momentum. 
Triangles and the fitted line $-0.743 \log qa+0.576$ are in the $\log U$, and diamonds and the fitted line $-0.747 \log qa+0.382$ are in the $U$-linear.}
\label{gl3.3}
\end{center}
\end{figure}

$\log-\log$ plot of ghost propagator is given in Fig 2. The difference of
 case 1 and case 3 is the global scale factor as in the case of gluon propagator. Case 2 is almost same as case 1.

\begin{figure}[htb]
\begin{center}
\epsfysize=110pt\epsfbox{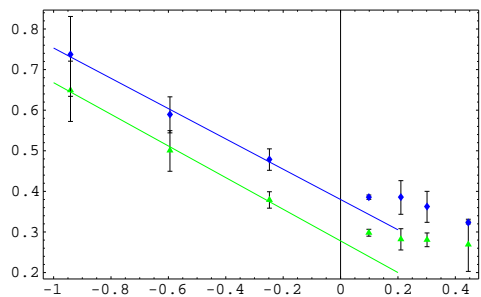}
\caption{The $\log$ of the unrenormalized ghost propagator multiplied by $(qa)^2$ as a function of the $\log qa$ where 
$q$ is the lattice momentum. 
Triangles and diamonds are the same as Fig\ref{gl3.3}.
The fitted line is $-0.390 \log qa+0.278$ in the $\log U$ and
$-0.373 \log qa+0.380$ in the $U$-linear.}
\end{center}
\end{figure}

This work was supported by KEK Supercomputer Project(No.00-57), and JSPS, Grant-in-aid for Scientific Research(C) (No.11640251).


\begin{thebibliography}{9}
\bibitem{Gv} V.N. Gribov, \NPB {\bf 139}, 1(1978).
\bibitem{KO} T. Kugo and I. Ojima, {\em Prog. Theor. Phys.} Supp. {\bf 66}, 1 (1979).
\bibitem{Zw} D. Zwanziger, %Nucl. Phys. {\bf B364}, (1991) 127,
 \NPB {\bf 412}, 657 (1994).
\bibitem{HdF} J.E. Hetrick and P.H. de Forcrand, \NPB
(Proc. Suppl.) {\bf 63}A-C, 838 (1998).
\bibitem{Fuj} K. Fujikawa, {\em Prog. Theor. Phys.} {\bf 61}, 627 (1979).
\bibitem{Hir} P. Hirschfeld, \NPB {\bf 157}, 37 (1979).
\bibitem{MO} J.E. Mandula and M. Ogilvie, \PLB {\bf 185}, 127 (1987).
\bibitem{NFY} H.Nakajima, S.Furui, A.Yamaguchi, ICHEP\\2000 contribution
paper, hep-lat/0007001.
\bibitem{Bny} V.G. Bornyakov, D.A. Komarov, M.I. Polikarpov, hep-lat/0009035.
\bibitem{NF} H.Nakajima and S. Furui, \NPB (Proc Suppl.){\bf 63}A-C, 635,865 (1999);
idem, \NPB (Proc Suppl.){\bf 83-84}, 521 (2000);  
Confinement III proc., hep-lat/9809078; %Lattice 99 proc., hep-lat/9909008;
QNP2000 proc., hep-lat/0004023.
\bibitem{Gius} L. Giusti, M.L. Paciello, S. Petrarca, B. Taglienti and 
M. Testa, hep-lat/9803021.
\bibitem{STR}The value $e/4$ in $\log U$ reported elsewhere so far was
 erroneous.
\end{thebibliography}
\end{document}